# Do we have privacy in the digital world?

**Kaveh Bakhtiyari**

Interactive Systems, Department of Computer & Cognitive Science
Faculty of Engineering, University of Duisburg-Essen
47057 Duisburg, North Rhine-Westphalia (NRW), Germany

Department of Electrical, Electronics and Systems Engineering,
Faculty of Engineering and Built Environment,
Universiti Kebangsaan Malaysia (The National University of Malaysia)
43600 UKM Bangi, Selangor Darul Ehsan, Malaysia

***Abstract** – Not really.*

When the computer was invented, it was a great idea to calculate and store data. Internet complemented the greatness of this technology. The hybrid creature of digital gadgets and the Internet is becoming smarter and more versatile every day in our daily life. Therefore, we are always concerned about our privacy. Personal privacy is a strong sense of human being. Wearing clothes, erecting walls, fence, partitions, and keeping distance are all the samples of privacy protection. But how much privacy do we have? Do the technology improvements, privacy statements and research progress guarantee our privacy? Let's see whether we have such barriers to protect our privacy in digital World.

Nowadays, we can find *privacy policy statements* on every famous or even infamous website which collects the users' information. This statement declares clearly the company's responsibility against the user's personal information. Usually, the most important section refers to keeping all the personal information safe and not to reveal them to any third party. But does it mean privacy to us?

Some computer users are scared of computer webcams (cameras). They even do not feel safe by sitting straight forward in front of a turned off the camera. Sometimes they cover it to feel more secure. The feeling of being watched and monitored is an important issue which has been shown in a lot of horror, action movies, and TV series. Whereas being watched is not only possible by cameras. There are other means of watching the users which are hidden from user's vision. Social networks, shopping websites, simple blogging websites, mice, keyboards, monitors, touch-screens, microphones, all of them can watch us instantly.

If you are an *Internet-heavy user*, you should be a member of at least one social network. People usually refer to social networks as a privacy disturbance media. The users set a bunch of settings on their social network profiles to limit the visibility of contents. By this way, they try to protect their privacy. Even though all those website settings improve the user's confidentiality, the user's posted information should be more concerned. Almost most of the Internet users are infringing their own privacies implicitly. Much information can be revealed to you which you may not want to say explicitly, regardless of how strong your privacy settings are. But let's see HOW?

Social network users reveal their information by posting on their profiles. This information can be personal or even general knowledge. People comment on posts, show their interest by liking (clicking on Like button) the items, and interact with their friends online. Over time, all these actions and interactions build a profile of preferences for each user. Data mining techniques can be employed to build a knowledge on the information on the user's profile. Then this profile can obviously say what this person thinks, what he/she likes, what he/she prefers, what he/she is interested in more and so on [1, 2].

Tag is another useful and privacy threaten invention. They are getting more popular on the social networks, weblogs, and websites. On textual posts, photos and videos, there is an ability to tag your friends and to be tagged by friends. Once a person is tagged, his/her profile would be linked to that post and related implicit information would be mined into the user's profile. It says where the user was at when, whom the user was with, what was the user feeling, who else has the same characteristics and preferences as this user [3, 4].

Still, there are some users that they do not post anything, they do not hit the LIKE or FOLLOW button, and they do not leave any comment on any other post. However, they have some friends listed in their friends' lists, they have some pages followed and liked to be followed in their accounts. They also have provided some basic information like education, birthday, languages, family, etc. in their profiles. According to the collaborative techniques, analyzing of a user's friends can result in understanding the user's preferences [5]. The preferences of the user's friends can expose the possible characteristics and favorites. The more active your friends are, the more information can be mined about you.

By not even interacting with social networks, still, it is easy to identify which posts and items the user is interested in. User's interaction with the website shows the feelings. There have been quite a long literature on human feelings recognition in human-computer interaction. Have you ever noticed why the social network feeds are usually shown on a vertical basis? This is not happening by chance. The pause time that you make in scrolling shows that you are more interested in that particular post or that specific person who has posted. Not only pauses, but also your mouse movements can disclose how you feel and think if you are tempted to comment or like a post, even though you do not touch any item [1, 2]. Now you may think that you have not joined any social network, so you are safe. Let's say you are a *medium Internet user*, thus you should probably communicate with other Internet users. You may use text, voice, and video in chats, online conference sessions and emails. These features are usually provided by a company for free or on a monthly payment basis. The law has allowed some governments (e.g. United Kingdom) to monitor these communication channels. Assuming the old eavesdropping method of having people sat and listen to the conversations is far beyond the imagination that everybody can believe it today. Artificial Intelligence (AI) is now in charge. Natural Language Processing (NLP) can analyze the user's text and extract the required information. Sound Signal Processing can analyze the human voice. Image Processing (IP) techniques can extract the features and interpret the information from the images and videos. This information can say how the user is feeling, how important or suspicious a conversation might be, and what the key points are in their conversations [6, 7].

Well, what if you have a *limited use of the Internet*? You may only use the Internet to read the news, pay the bills and do some daily shopping. Have you ever noticed that some famous shopping websites recommend you the items, as they may know who you are and what you may like? Recommender Systems are widely used in shopping websites. When a user buys a single item, the system would automatically evaluate multiple criteria such as similar users' behavior, the bought item features, geographical and personal features to come up with new recommendations. It means that the shopping websites do not need us to buy tons of items to learn our preferences [5]. There is some literature on recommender systems to predict the relevant items to the first-time users.

Besides that, website cookies have a history of discussion on privacy. Cookies are tiny files which a website stores on our computer. Each time we return, the stored information can be read. The cookies are very important in tracking our browsing, signing in, remembering the password, etc. Cookies can also be transferred from one website to another, therefore we can be tracked from one website to another website. When we return to a website or browse a website for the first time, the website may greet us. That shows we have been tracked. Deleting and wiping the computer from cookies is not even a good solution to avoid from being tracked. The network IP address is also used for tracking the geographical location. Although IP address is not usually fixed all the time, they can reveal information on where the person is located now.

Internet browsing basically includes user's interaction, mouse movements, keyboard keystrokes, and cookie tracking. This information is usually gathered and analyzed by corporations to improve their production. For example, the heat map is a sample which can be used to interpret the user's interactions. These interpretations can say what you feel, what you like, and what you are looking for [1, 2].

I believe the big eyes surrounding us as smartphones, tablets, and PDAs are becoming more popular and more users are joining the club of gadget lovers every day. All these gadgets connect to the Internet, they sync the information with the Internet servers. They have all users' main profiles connected together. Smart gadgets (including computers) which have processors and memory are really smart. They are alive. They know about your Facebook account, your twitter, your emails, your photos, your cloud files & documents, your current position (GPS, A-GPS), your contacts, the people you contact with more by different means and applications, and they know a lot more than we imagine. The smarter the gadget is, the more information and data can be revealed about a user. There have been some reports that even the mobile phones can be used to eavesdrop the user's environment by the governments even when the phone is switched off.

People sometimes ask: Why do they need this information? They think that only famous people are tracked and watched. Maybe the personal information of a single person is not very important except for marketing purposes. However, according to the prediction and classification methods of data mining, prediction of the future is possible by analyzing the data. With the amount of data which already exists in social networks (such as Facebook), it is nearly possible to predict the near future of every single person by considering his/her timeline and classifying the similar users through collaborative and content-based techniques. This achievement is more incredibly powerful than we can imagine. Two years ago, Onion News Network announced a fake news that CIA has more funding for the Facebook program as an information gathering system. Also, Onion News Network announced that Christopher Sartinsky (a cinematographer, director, and editor), deputy CIA director (a fake position), said about Facebook: "It is a dream come true for the CIA.[1]" Even though it was a fake news on YouTube, but it is true that the data on Facebook or any other social networks and/or email providers are extremely useful and valuable for intelligence agencies. As mentioned, it is not all about Facebook or social networks only, but it is all about the Internet and our digital life.

Being connected to the Internet and having a smartphone in your pocket is equal to being watched with no privacy regardless of what the privacy statements say. Maybe using the stupid phones (old fashioned phones) and computers off the Internet would be a better option for those who concern more about privacy. In addition, for those people who are very strict in their own privacy, being away from all digital gadgets and keeping distance with all people who use these devices are the only possible solutions to protect their privacy.

However, a question remains. Should we accept the digital life with no or less privacy? Alternatively, should we keep our privacy by being off the technology?

---

[1] http://www.youtube.com/watch?v=8G28TTRjQV0